\documentstyle[12pt]{article}
\catcode`\@=11

\def\@normalsize{\@setsize\normalsize{15pt}\xiipt\@xiipt
           \abovedisplayskip 14pt plus3pt minus3pt%
           \belowdisplayskip \abovedisplayskip
           \abovedisplayshortskip  \z@ plus3pt%
           \belowdisplayshortskip  7pt plus3.5pt minus0pt}

\def\small{\@setsize\small{13.6pt}\xipt\@xipt
           \abovedisplayskip 16pt plus3pt minus3pt%
           \belowdisplayskip \abovedisplayskip
           \abovedisplayshortskip  \z@ plus3pt%
           \belowdisplayshortskip  7pt plus3.5pt minus0pt
             \def\@listi{\parsep 4.5pt plus 2pt minus 1pt
             \itemsep \parsep
             \topsep 9pt plus 3pt minus 3pt}}

\def\underline#1{\relax\ifmmode\@@underline#1\else
        $\@@underline{\hbox{#1}}$\relax\fi}
\@twosidetrue

\long\def\@makecaption#1#2{
 \vskip 10pt
 \setbox\@tempboxa\hbox{#1: #2}
 \ifdim \wd\@tempboxa >\hsize #1: #2\par \else \hbox
        to\hsize{\box\@tempboxa\hfil}
 \fi}

\catcode`\@=12

\relax

\newskip\humongous \humongous=0pt plus 1000pt minus 1000pt

\newif\ifdtup

\def\oldreffmt#1{\rlap{[#1]} \hbox to 2\parindent{}}

\def\figfmt#1{\rlap{Figure {#1}} \hbox to 1in{}}

\def\slash#1{#1\!\!\!/\!\,\,}
\def\beq{\begin{equation}}
\def\eeq{\end{equation}}
\def\bea{\begin{eqnarray}}
\def\eea{\end{eqnarray}}

\def\bq{\begin{quote}}
\def\eq{\end{quote}}

\relax

\def\SM{Standard Model }

\def\EW{electro--weak }

\def\GB{Goldstone Boson }

\def\GBs{Goldstone Bosons }

\def\bar{\overline}
\def\phi{\varphi}

\def\addcontentsline#1#2#3{}

\begin{document}


{    
\def\thefootnote{\fnsymbol{footnote}}
\thispagestyle{empty}

\ \vskip -.8cm
\ \hskip 12.1cm HD--THEP--93--19

\ \vskip -1.2cm

\vskip 2.3cm
\begin{center}
      {\Large\sc\bf The Phenomenological Viability of} \\
      \ \\
      {\Large\sc\bf Top Condensation}\\
\vskip 1.7cm
      {\sc Manfred Lindner\footnote{Heisenberg Fellow}}\\

\vskip .8cm
      {\sl  Institut f\"ur Theoretische Physik\\
      der Universit\"at Heidelberg\\
      Philosophenweg 16, D--W--6900 Heidelberg}\\
\end{center}

{\def\thefootnote{}
 \footnote{Email: Y29 at VM.URZ.UNI--HEIDELBERG.DE}
}

\vskip 2.6cm
   \begin{center}{\Large\bf Abstract}\end{center}
\par \vskip .05in
We discuss how the full dynamics of top condensation models can
modify the relations between the physical top mass, the amount
of custodial $SU(2)$ violation and the weak gauge boson masses.
It is emphasized that it is possible to get phenomenologically
acceptable relations between $\Delta\rho$, $m_t$ and $M_W$ and
that in addition the scale of new physics can be chosen to be
${\cal O}(TeV)$ such that a fine--tuning problem is avoided.
} 

\vfill
\noindent
{\sl To appear in the Proceedings of the 4th Hellenic School
     on Elementary Particle Physics}

\newpage

\setcounter{page}{1}
\setcounter{footnote}{0}

The formation of a $\bar t t$ condensate by some ``pairing force'' could be
responsible for a dynamical breaking of the \EW symmetry \cite{BHL,Hab}.
A theory creating such a condensate would naturally explain a heavy
top mass, would be very helpful to avoid Flavour Changing Neutral Current
(FCNC) problems  and would be very attractive due to its economy.
Simple initial realizations of top condensation were based on effective
Nambu--Jona-Lasinio (NJL) models with non--renormalizable
four--fermion interactions. This led however to discussions about higher
dimensional operators \cite{Hase} which depend crucially on how
``effective'' or ``fundamental'' the four--fermion interactions are.
Subsequently fundamental four--fermion theories where proposed
\cite{L4f} while other authors justify an effective NJL description
of strongly coupled broken gauge theories \cite{TOPCOLOR,U1U1,King,Bon,Mar}.

Independently of such questions we study here the phenomenological
viability of top condensation ideas by assuming essentially only
that we know the solution $\Sigma_t(p^2)$ of some relevant Schwinger--Dyson
(``gap'') equation for the dynamically generated top mass. Thus we
pretend to know the \EW symmetry breaking top propagator to be
\beq
S_t({\rm p}^2) = \frac{i}{\slash{{\rm p}} -\Sigma_t({\rm p}^2)}~,
\label{SF}
\eeq
with the (pole) top mass $m_t=\Sigma_t(m_t^2)$. All other quarks and
leptons are assumed to be massless. Without specifying the gap equation
we assume furthermore that for the theory under consideration\footnote{This
is e.g. justified for asymptotically free theories where chiral
symmetry breaking disappears as $p^2\rightarrow\infty$.}
$\Sigma_t(p^2)\stackrel{p^2\rightarrow\infty}{\bf\longrightarrow}0$
and that there is only one unique solution for $m_t$.

The breaking of the \EW symmetry (i.e. $\Sigma_t\neq 0$) is assumed
to be the result of unspecified new strong forces acting only on the
known quarks and leptons and especially on the $t-b$ doublet.
The emergence of a top condensate breaks global symmetries and the
resulting \GBs are ``eaten'' in a dynamical Higgs mechanism such that
$W$ and $Z$ become massive. Presumably such a theory does not change
significantly if the weak $U(1)_Y$ coupling $g_1$ is sent to
zero\footnote{Indirectly (via vacuum alignment) a small $U(1)_Y$ coupling
could be very important such that $g_1=0$ should be understood as the result
of the limiting procedure $g_1\rightarrow 0$.}. In the limit $g_1=0$ the
corrections which give mass to the $W_3$ and $W_\pm$ propagators must
be induced by the fermions which are representations under both
$SU(2)_L$ and the new strong force. We should therefore study the
contributions of $\Sigma_t$ to the vacuum polarizations of the $W$
and $Z$ propagators. In an expansion in powers of $g_2^2$ the leading
contribution is given by diagrams which connect the $W_\pm$ or $W_3$
line to a fermion pair from both sides. There are two ways how these
four internal fermion lines can be connected: By inserting twice the full
fermionic propagators or by inserting once the full four--fermion
Kernel of the new, strong interaction. Note that in leading order
$g_2^2$ the fermion propagators and the Kernel do not contain any
\EW gauge boson propagation themself since this would cost at least an
extra power of $g_2^2$. Insertions of fermionic vacuum polarizations
into higher order \EW loop diagrams, for example, are suppressed
by corresponding powers of $g_2^2$. Thus in leading order $g_2^2$, but
exact in the new strong coupling, the $W$ propagator is graphically
represented by Fig.~\ref{F1}. The first contribution is a generalization
of the leading \SM diagrams with hard fermion masses replaced by $\Sigma$'s,
i.e. the sum of all one particle irreducible diagrams which contribute
to the dynamically generated fermion masses. The second contribution
contains the exact Kernel $K$ of the strong forces
responsible for condensation and it is useless to expand this Kernel
perturbatively in powers of the coupling constant of the new strong force.
The Goldstone theorem tells us however that the Kernel must contain poles
of massless \GBs due to the breaking of global symmetries by the fermionic
condensates. This is symbolically expressed by the second line of
Fig.~\ref{F1}, where $\tilde K$ does not contain any further poles of
massless particles. But $\tilde K$ may (and typically will) contain
all sort of massive bound states which could e.g. be vectors, Higgs--like
scalars etc. in all possible channels.

The \GB contributions\footnote{Which are essential for a gauge
invariant dynamical Higgs mechanism.} shown in the second line of
Fig.~\ref{F1} were used by Pagels and Stokar \cite{PaSto} to obtain
a relation between the $\Sigma$'s and the \GB decay constants.
Their derivation uses the fact that only the
\GBs contribute a term proportional $p_\mu p_\nu/p^2$ to the $W$
polarization at vanishing external momentum, but this method ignores possible
contributions from $\tilde K$ which enter indirectly via the use of Ward
identities. The $p_\mu p_\nu/p^2$ contributions to $\Pi_{\mu\nu}$ are
balanced (up to small corrections from $\tilde K$) by $g_{\mu\nu}$ terms
created by the first diagram on the {\em rhs} of Fig.~\ref{F1}.
Following ref.~\cite{BluLi} we derive a relation between the $\Sigma$'s
and the \GB decay constants from these $g_{\mu\nu}$ terms. The result
can be compared with the Pagels--Stokar relation and we will see that
contributions from $\tilde K$ are significantly suppressed. Let us
therefore work with rescaled fields such that gauge couplings appear
in the kinetic terms of the gauge boson Lagrangian like
$\left(-1/4g^2\right)\left(W_{\mu\nu}\right)^2$.
Since we do not include any propagating $W$ bosons we need not gauge fix
at this stage. The inverse $W$ propagator can be written as
\beq
\frac{1}{g_2^2}D_{W,\mu\nu}^{-1}(p^2) = \frac{1}{g_2^2}
  \left(-g_{\mu\nu}+\frac{p_\mu p_\nu}{p^2}\right)~p^2
- \Pi_{\mu\nu}(p^2)~,
\label{invprop}
\eeq
with the polarization tensor
$\Pi_{\mu\nu}(p^2)=\left(-g_{\mu\nu}p^2+p_\mu p_\nu\right)\Pi(p^2)$.
At vanishing external momentum the first fermion loop on the {\em rhs}
of Fig.~\ref{F1} contributes to $\Pi_{\mu\nu}$
\beq
\Pi_{\mu\nu} = -iZ^2N_c~\int\frac{d^4k}{(2\pi)^4}~
\frac{
{\sl Tr}\left[\Gamma_\mu(\slash k+\Sigma_1(k))\Gamma_\nu
(\slash k+\Sigma_2(k))\right]
}
{(k^2-\Sigma_1(k)^2)(k^2-\Sigma_2(k)^2)}~,
\label{Pimunu}
\eeq
where $N_c$ is the number of colors, $Z^{-1}=\sqrt{2}, 2$
in the charged and neutral channel, respectively,
$\Gamma_\alpha=\left(\frac{1-\gamma_5}{2}\right)\gamma_\alpha$, and
${\bf +i\epsilon}$ is generally implied in the denominator.
In the neutral channel we get corrections from $\bar t t$
(i.e. $\Sigma_1=\Sigma_2=\Sigma_t$), $\bar b b$
(i.e. $\Sigma_1=\Sigma_2=\Sigma_b\equiv 0$) and in the charged channel
contributes only $\bar t b$ or $\bar b t$
(i.e. $\Sigma_1=\Sigma_t$, $\Sigma_2=\Sigma_b\equiv 0$).
By naive power counting eq.~(\ref{Pimunu}) has quadratic and logarithmic
divergences. Since we assumed
$\Sigma_i(p^2)\stackrel{p^2\rightarrow\infty}{\bf\longrightarrow}0$
for the top quark and all other fermions
we find that the divergences of $\Pi_{\mu\nu}(p^2)$ are identical to those
calculated for $\Sigma_i\equiv0$. It makes therefore sense to split
$\Pi_{\mu\nu}(p^2)=\Pi^0_{\mu\nu}(p^2)+\Delta\Pi_{\mu\nu}(p^2)$ where
$\Pi^0_{\mu\nu}$ is defined as $\Pi_{\mu\nu}$ for $\Sigma_i\equiv0$.
$\Pi^0_{\mu\nu}$ is then an uninteresting $\Sigma_i$ independent constant
which contains all divergences and needs renormalization. Contrary the
interesting $\Sigma_i$ dependent piece
$\Delta\Pi_{\mu\nu}=\Pi_{\mu\nu}-\Pi^0_{\mu\nu}$
is finite, even when the external momentum is sent to zero. Thus
\bea
\Delta\Pi_{\mu\nu} &=&
  -iZ^2N_c~\int\frac{d^4k}{(2\pi )^4}~\left\{
    \frac{
    {\sl Tr}
    \left[\Gamma_\mu(\slash{k}+\Sigma_1)\Gamma_\nu(\slash{k}+\Sigma_2)\right]
           }{(k^2-\Sigma_1^2)(k^2-\Sigma_2^2)}
  -
    \frac{
    {\sl Tr}\left[\Gamma_\mu\slash{k}\Gamma_\nu\slash{k}\right]
           }{k^4}
                                 \right\} \\
                  &=&
  - g_{\mu\nu}~\frac{Z^2N_c}{(4\pi)^2} \int\limits_0^\infty dk^2~
    \frac{k^2(\Sigma_1^2+\Sigma_2^2) - \Sigma_1^2\Sigma_2^2}
    {(k^2-\Sigma_1^2)(k^2-\Sigma_2^2)}~,                 \label{finalDPi}
\eea
where angular integration was performed in Euclidean space and subsequently
continued back to Minkowski space. Under the integral one has as usual
${\sl Tr}\left[\Gamma_\mu\slash{k}\Gamma_\nu\slash{k}\right]=-g_{\mu\nu}k^2$
and ${\sl Tr}\left[\Gamma_\mu\Gamma_\nu\right] = 0$. Note that this
separation procedure for $\Delta\Pi_{\mu\nu}$ does not spoil gauge
invariance.

The \GB decay constants $F_i^2$ are the poles of $\Pi(p^2)$ at vanishing
external momentum. For our definition of $\Pi_{\mu\nu}$ we find that
$F_i^2$ is identical to eq.~(\ref{finalDPi}) without the factor
$-g_{\mu\nu}$. Using $Z$ for the charged and neutral channel one finds
\beq
F_\pm^2 = \frac{N_c}{32\pi^2}\int\limits_0^\infty dk^2~
          \frac{\Sigma_t^2}{k^2-\Sigma_t^2}~,\quad
F_3^2   = \frac{N_c}{32\pi^2}\int\limits_0^\infty dk^2~
          \frac{k^2\Sigma_t^2-\frac{1}{2}\Sigma_t^4}{(k^2-\Sigma_t^2)^2}~,
\label{FF}
\eeq
such that
\beq
F_3^2-F_\pm^2 = \frac{N_c}{64\pi^2}\int\limits_0^\infty dk^2~
                \frac{\Sigma_t^4}{(k^2-\Sigma_t^2)^2}~.
\label{finaldF2}
\eeq
Eq.~(\ref{FF}) for $F_\pm^2$ is equivalent to the result obtained by Pagels
and Stokar \cite{PaSto} from the $q_\mu q_\nu/q^2$ contributions of \GBs
to $\Pi_{\mu\nu}$. The result for the neutral channel in eq.~(\ref{FF})
looks however somewhat different. But by using the integral identity
\beq
\int\limits_0^\infty dx ~\frac{x^2f(x)^\prime - f(x)^2}
{\left( x-f(x) \right)^2} = f(\infty )~,
\label{intid}
\eeq
for $x=k^2$ and $f=\Sigma_t^2$ we can rewrite eq.~(\ref{FF}) into
\beq
F_3^2 = \frac{N_c}{32\pi^2}\int\limits_0^\infty dk^2~k^2~
        \frac{\Sigma_t^2- k^2\Sigma_t\Sigma_t^\prime}{(k^2-\Sigma_t^2)^2}~,
\label{PASTOres}
\eeq
where $\Sigma_t^\prime = d\Sigma_t /dk^2$. Even though this looks now
formally similar to the Pagels--Stokar result it differs by a factor 2
in front of the derivative term in the nominator of eq.~(\ref{PASTOres}).
This difference may appear less important, but we will see
that in the limit of a hard top mass our method produces the correct
$\rho$--parameter, while the Pagels--Stokar result produces 3/2 times the
correct answer. Additionally our expression leads also to a better numerical
estimate of $f_\pi$ if we follow the methods of ref.~\cite{PaSto}.

The $\rho$--parameter \cite{rhoinvent} is defined as $\rho:=F_\pm^2/F_3^2$
which can now be written as
\beq
\rho = 1+\Delta\rho =
\frac{F_\pm^2}{F_3^2} = \left(1+\frac{(F_3^2-F_\pm^2)}{F_\pm^2}\right)^{-1}
\simeq 1 - 2~\frac{(F_3^2-F_\pm^2)}{v^2}~,
\label{rhoaprox}
\eeq
and from eq.~(\ref{finaldF2}) we find the contribution of the $t-b$
doublet
\beq
\Delta\rho = \frac{-N_c}{32\pi^2v^2}~\int\limits_0^\infty dk^2~
             \frac{\Sigma_t^4}{(k^2-\Sigma_t^2)^2}~,
\label{finalrho}
\eeq
where we used $F_\pm^2=v^2/2$ with $v\simeq 175~GeV$ in the denominator.
Model independent parametrizations of radiative corrections parametrize
the information contained in $\Delta\rho$ essentially in the variables
$T$ \cite{PesTa} or $\epsilon_1$ \cite{Alta}.

With the expressions for $\Delta\rho$ in eq.~(\ref{finalrho}) and
$F_i^2$ in eq.~(\ref{FF}) we can calculate for given
$\Sigma_t(p^2)\stackrel{p^2\rightarrow \infty}{\longrightarrow}0$ three
independent observable quantities which are one of the weak gauge boson
masses (either $M_W^2 = g_2^2F_\pm^2$ or $M_Z^2 = (g_1^2+g_2^2)F_3^2~$),
$\Delta\rho$ and furthermore the physical top mass $m_t$. These three
quantities are dominated by different momenta and therefore
$\Sigma\neq constant$ leads to a different relation than a constant, i.e.
hard mass. It is instructive to look at the degree of convergence of the
involved integrals. The \GB decay constants $F_i^2$ are formally log.
divergent, but are finite with our assumption on $\Sigma_t(p^2)$. In that
case renormalization is not needed, but due to the formal log. divergence
$\Sigma_t$ contributes with equal weight at all momentum scales. In other
words, the magnitude of $F_i^2$ depends crucially on the high energy tail
of $\Sigma_t$. The difference $F_\pm^2 - F_3^2$ has better convergence
properties and is always finite, even for $\Sigma_t(p^2)=constant$. This
implies that $\Delta\rho$ is finite, as it should be, and it is most
sensitive to infrared scales somewhat above $m_t$. Finally $m_t$ is of
course only sensitive to one point, namely $m_t=\Sigma_t(m_t^2)$.

We would like to study now corrections in the relation between $m_t$,
$M_W$ and $\Delta\rho$ when $\Sigma_t$ is the solution of a hypothetical
Schwinger--Dyson equation which deviates from $\Sigma_t=m_t= constant$.
First we would like to see if the correct \SM result emerges for a $t-b$
doublet when $\Sigma_t\rightarrow m_t$. Therefore we set
\beq
\Sigma_t(p^2) = m_t\,\Theta(\Lambda^2-p^2)~,
\label{topan1}
\eeq
and ignore again the $b$ quark mass. From eq.~(\ref{finalrho}) we obtain
for our ansatz
\beq
\Delta\rho
= \frac{N_cm_t^2}{32\pi^2v^2}\left(\frac{1}{1-m_t^2/\Lambda^2}\right)
  ~\stackrel{\Lambda\rightarrow\infty}{\longrightarrow}~\Delta\rho^{SM}
= \frac{N_c\alpha_{em}}{16\pi\sin^2\theta_W\cos^2\theta_W}
  ~\frac{m_t^2}{M_Z^2}~.
\label{rhotheta}
\eeq
Note that in the limit $\Lambda\rightarrow\infty$ (i.e. a hard, constant
top mass) we obtain correctly the leading \SM value while the Pagels--Stokar
relation would produce 3/2 times the \SM result.
For finite $\Lambda$ eq.~(\ref{rhotheta}) describes furthermore the
modification of the \SM result due to a physical high energy momentum
cutoff. Such a cutoff makes $\Delta\rho$ a little bit more positive than
in the \SM which implies for a fixed experimental value of $\Delta\rho$
a lower top mass prediction. From eq.~(\ref{FF}) it is in addition
possible to determine $M_W$ for the ansatz eq.~(\ref{topan1})
\beq
M_W^2
 = g_2^2F_\pm^2
 = \frac{g_2^2N_c}{32\pi^2}
         ~\int\limits_0^{\Lambda^2}dk^2~\frac{\Sigma_t^2}{k^2-\Sigma_t^2}
 =\frac{g_2^2N_c}{32\pi^2}~m_t^2~
        \ln{\left(\frac{\Lambda^2-m_t^2}{m_t^2}\right)}~.
\label{MWmttheta}
\eeq
Taking as experimental input $M_W=80.14\pm 0.27~GeV$,
$\Delta\rho = 0.005\pm 0.008$, $\alpha_{em}^{-1}(M_Z^2) = 127.8\pm 0.1$
and $\sin^2\theta_W^{eff}(M_Z^2) = 0.2318\pm 0.0007$ we plot in
Fig.~\ref{F2} the two central top mass values resulting from
eqs.~(\ref{rhotheta}) and (\ref{MWmttheta}) as a function of $\Lambda$
(dashed lines).

The ansatz eq.~(\ref{topan1}) can be viewed as the result of a
Nambu--Jona-Lasinio (NJL) gap equation of top condensation as for
example in the model of Bardeen, Hill and Lindner (BHL) \cite{BHL}.
In fact a NJL gap equation is the simplest conceivable Schwinger--Dyson
equation where $\Sigma_t$ is forced to be a constant. Fig.~\ref{F2} shows
clearly that ultra high values of $\Lambda$ and the experimental errors are
required to get the two top mass values in agreement. For such high
$\Lambda$ the effective Lagrangian is valid for many orders of magnitude
which led in the BHL analysis to the so--called ``renormalization group
improvement''. This means in the current language
that $\Sigma_t=constant$ is replaced by
$\Sigma_t=g_t(p^2)v$, where $v=175~GeV$ and $g_t(p^2)$ is the solution
of the one--loop renormalization group equation. In BHL the predicted
top mass is then the ``effective fixedpoint'' of the renormalization
group flow. The same result could be seen in eq.~(\ref{MWmttheta}) since
the effective fixedpoint dictates the shape of $\Sigma(p^2)$ for many
orders of magnitude. The BHL scenario has however phenomenological problems.
First the very high value of $\Lambda$ is nothing else then the old
hierarchy problem which appears now as a fine--tuning of the four--fermion
coupling $G$. Furthermore the infrared fixedpoint prediction is higher
than the dashed curve resulting from eq.~(\ref{MWmttheta}) which is
shown in Fig.~\ref{F2} and has (within newest experimental errors) no
intersection with the line resulting from eq.~(\ref{rhotheta}). Thus this
simplest scenario seems unacceptable even for very high values of $\Lambda$.

Remembering that $\Delta\rho$ and $M_W$ are sensitive to details
of $\Sigma_t$ in a different way we should ask ourselfs if the above
problems can be solved by modifications of the solution $\Sigma_t(p^2)$.
The answer is of course yes, and we illustrate now the two most
important type of changes: The addition of a slowly falling tail
and/or the addition of a ``bump'' somewhat above $m_t$.

First we consider a very rough ansatz for a ``bump'' between
$\Lambda_1$ and $\Lambda$ with $m_t<\Lambda_1<\Lambda$
by modifying eq.~(\ref{topan1})
\beq
\Sigma_t(p^2) = \left\{
\begin{array}{ll}
0                 & {\rm for~} p^2 > \Lambda^2~;\\
\sqrt{r}\cdot m_t & {\rm for~} \Lambda_1^2 \leq p^2 \leq \Lambda^2~;\\
m_t               & {\rm for~} p^2<\Lambda_1^2~,
\end{array}\right.
\label{topan2}
\eeq
where $\Sigma$ is changed between $\Lambda_1$ and $\Lambda$. For $r>1$
there is an extra ``bump'' between $\Lambda_1$ and $\Lambda$ which affects
$\Delta\rho$. For $\Lambda^2,\Lambda_1^2 \gg m_t^2,rm_t^2$ we get
\beq
\Delta\rho \simeq
\frac{N_cm_t^2}{32\pi^2v^2}\left(1+\frac{m_t^2}{\Lambda^2}
        -\left[\frac{m_t^2(\Lambda^2-\Lambda_1^2)}
         {\Lambda^2\Lambda_1^2}~(r^2-1)\right]
\right)~,
\label{rhobump}
\eeq
where extra contributions due to $r\neq 1$ and $\Lambda_1\neq\Lambda$
are isolated in square brackets. We can see that the bump counteracts
the effect of the cutoff and makes $\Delta\rho$ less positive. In
principle the bump can even be chosen to make $\Delta\rho$ vanish.
The relation eq.~(\ref{MWmttheta}) between $m_t$ and $M_W$ becomes also
modified. For $\Lambda^2,\Lambda_1^2 \gg m_t^2,rm_t^2$ we get approximately
\beq
M_W^2 \simeq \frac{g_2^2N_c}{32\pi^2}~m_t^2~
\left(\ln{\left(\frac{\Lambda^2-m_t^2}{m_t^2}\right)}
        +\left[(r-1)\ln{\left(\frac{\Lambda^2}{\Lambda_1^2}\right)}\right]
\right)~,
\label{MWmtbump}
\eeq
where extra contributions due to the bump are again isolated in square
brackets.

Now we add a slowly falling high energy tail to the last ansatz
eq.~(\ref{topan2})
\beq
\Sigma_t(p^2) = \left\{
\begin{array}{ll}
{\rm equation~(\ref{topan2})}                   &{\rm for~}p^2<\Lambda^2~;\\
\sqrt{r}
m_t~\left(\frac{p^2}{\Lambda^2}\right)^{-\alpha}&{\rm for~}p^2>\Lambda^2~,
\end{array}\right.
\label{topan3}
\eeq
where $\alpha>0$ is assumed. This high energy tail which is parametrized
by $\alpha$ leads to
\beq
\Delta\rho \simeq
\frac{N_cm_t^2}{32\pi^2v^2}\left(1+\frac{m_t^2}{\Lambda^2}
        -\left[\frac{m_t^2(\Lambda^2-\Lambda_1^2)}
         {\Lambda^2\Lambda_1^2}~(r^2-1)\right]
        -\left\{\frac{r^2}{4\alpha+1}~\frac{m_t^2}{\Lambda^2}\right\}
\right)~,
\label{rhotail}
\eeq
and
\beq
M_W^2 =\frac{g_2^2N_c}{32\pi^2}~m_t^2~
\left(\ln{\left(\frac{\Lambda^2-m_t^2}{m_t^2}\right)}
        +\left[(r-1)\ln{\left(\frac{\Lambda^2}{\Lambda_1^2}\right)}\right]
        +\left\{\frac{r}{2\alpha}\right\}\right)~,
\label{MWmttail}
\eeq
where the extra corrections due to the tail are isolated in curly
brackets.

Note that we are looking for a scenario which simultaneously avoids
the fine--tuning problem and which is phenomenologically acceptable.
Consequently $\Lambda$ and $\Lambda_1$ should be $TeV$--ish and
the top mass values required from the $\Delta\rho$-- and $M_W$--data
should agree. This requires consequently some gap equation with a
generic condensation scale ${\cal O}(TeV)$ capable of producing a bump,
and a tail -- maybe of the type discussed in ref.~\cite{BluLi}.
The asymptotic high energy behaviour
of $\Sigma_t$ might be described by a renormalization group equation
if the spectrum of the theory does not contain further mass thresholds.
This would imply a logarithmic tail and the parameter $\alpha$ should be
very small. We could for example fix $\alpha$ in the minimal scenario by
expanding the Higgs less one--loop renormalization group equation
for $g_t$ in the \SM. This would lead to $\alpha \simeq 0.04$.
For such small values of $\alpha$ the tail leads to mild effects
in the $\rho$--parameter and drastic changes in the $M_W$--$m_t$ relation.

We can illustrate the effects of the combined bump and tail
by plotting  eqs.~(\ref{rhotail}) and (\ref{MWmttail}) in Fig.~\ref{F2}
as solid lines for the parameters $r=2$, $\Lambda=2\Lambda_1$,
$\Lambda_1=2m_t$ and $\alpha=0.04$.
The small value of $\alpha$ (corresponding to a logarithmic high energy
tail of $\Sigma_t$) influences mostly the $M_W$--$m_t$ relation while
the bump affects essentially only the $\Delta\rho$--$m_t$ relation.
Taking into account experimental and theoretical errors the two top mass
values agree for low values of $\Lambda$ consistent with the above
assumptions and avoiding fine--tuning. We have thus illustrated that
structured solutions of $\Sigma_t$ can solve the fine--tuning problem,
i.e. allow for $\Lambda$--values within a few $TeV$. Furthermore the
predicted $m_t$--$M_W$--$\Delta\rho$ relations are modified to be
consistent with the data on $M_W$ and $\Delta\rho$. The predicted
top mass differs however typically somewhat from its \SM
value -- something that will only be tested by a direct search for the
top quark. A bump and a tail as discussed could for example be
relevant in models of top condensation where heavy gauge bosons trigger
condensation \cite{U1U1} or in bootstrap scenarios where the
$t$--channel effects of a composite Higgs are non--negligible \cite{Nambu}.

There are other \EW observables which are sensitive to the top mass value
like for example the $Z{\bar b}b$ vertex. If $m_t$ is replaced by
$\Sigma_t$ in the relevant diagrams then one finds however that the
top mass dependence is replaced by sensitivity to $\Sigma_t$ at
low momenta. Thus in a first approximation these quantities
depend essentially on the pole mass. There are however corrections
which should become observable if high enough precision can be reached.

In summary we find that top condensation models are both phenomenological
viable and natural if $\Sigma_t$ has suitable structure. The calculation
of $\Sigma_t$ from first principles is however in general very difficult
for a given model due to the non--perturbative nature of the relevant
Schwinger--Dyson equation. But a reliable probe of the discussed effects
will emerge when the Fermilab Collider starts to push the direct top
mass limits into the \SM window. If the above ideas are relevant then
the top quark should not be found precisely in the often cited \SM window
but somewhat higher.

%


\newpage

\ \vskip -1.5cm

\centerline{\huge\bf Figures}


\begin{figure}[ht]
\includegraphics{corfig1.psf}
\vspace*{0.8cm}
\unitlength1cm
    \begin{picture}(15,3.6)
    \put(7.95,2.75){$K$}
    \put(12.45,0.55){$\tilde K$}
    \put(7.95,1.2){$\frac{i}{q^2}$}
    \end{picture}
\caption{\label{F1}
{\sl The $W$ propagator in leading order $g_2^2$ and exact
in the new non--perturbative interactions. Fermionic self--energies
are represented as fat dots and the four--fermion Kernel $K$
is represented by a fat circle. In the second line the Kernel is
split into \GB contributions (which arise due to the broken global
symmetries with some non--trivial vertex function) and $\tilde K$
(which has no further massless poles).}}
\end{figure}


\begin{figure}[hb]
\includegraphics{corfig2.psf}
\vspace*{6.5cm}
\unitlength1cm
    \begin{picture}(15,4)
    \end{picture}
\caption{\label{F2}
{\sl The predicted (pole) top mass $m_t$ versus $\Lambda$
using $\Delta\rho$ and $M_W$ as experimental input. The upper dashed
line follows from eq.~(\ref{MWmttheta}) and the lower dashed line
from eq.~(\ref{rhotheta}). The solid lines follow from the combined
bump and tail ansatz for $\Sigma_t$ showing that low values of
$\Lambda$ are then possible.}}
\end{figure}

\end{document}